\documentstyle[sprocl]{article}

\input{psfig}

\def\be{\begin{equation}}
\def\ee{\end{equation}}
\def\bea{\begin{eqnarray}}
\def\eea{\end{eqnarray}}

\begin{document}

\title{Nuclear Corrections in Neutrino Deep Inelastic Scattering and 
the Extraction of the Strange Quark Distribution}


\author{C. BOROS} 

\address{Department of Physics and Mathematical Physics, and 
         Special Research Center for the Subatomic Structure of Matter, 
         University of Adelaide, Australia\\ 
         {\tt E-mail: cboros@physics.adelaide.edu.au}} 

\maketitle\abstracts{
Recent measurement of the structure function 
$F_2^\nu$ in neutrino deep inelastic scattering 
allows us to compare structure function measured in neutrino and charged 
lepton scattering for the first time with reasonable 
precision. The comparison  
between neutrino and muon structure functions 
made by the CCFR Collaboration 
indicates that there is a discrepancy 
between these structure functions 
at  small Bjorken x values. In this talk 
I examine two effects which might account for this 
experimental discrepancy: nuclear shadowing corrections 
for neutrinos and contributions from strange and anti strange quarks.}  

\section{Introduction} 
Recently, there has been much interest in the structure
function  $F_2^{\nu}$ measured  by the CCFR Collaboration
\cite{CCFR} in neutrino deep inelastic scattering.  
This measurement makes it possible to compare structure functions
extracted from neutrino-induced reactions
with those measured in charged lepton-induced ones, 
to test the universality of parton distribution
functions and to determine the strange quark density of the nucleon. 

The CCFR Collaboration compared the neutrino structure function
$F_2^\nu$ extracted from their data taken on an iron target \cite{CCFR}
with $F_2^\mu $ measured for the deuteron by the NMC
Collaboration \cite{NMC}. They found that, while there is a 
very good agreement between 
the two structure functions for intermediate values of
Bjorken $x$ ($0.1 \le x \le 0.4$), in the small
$x$-region ($x < 0.1$), the two structure functions differ 
by as much as 10-15$\%$.

However, there are many corrections which have to be applied to the 
data for such a comparison to be sensible.  
Since the data have been already corrected for charm 
threshold effects and non-isoscalarity 
of  the iron target, in this talk, I focus on two remaining important 
corrections: 

(i) The neutrino structure function is measured 
on an iron target and therefore it has to be corrected for nuclear effects. 
Nuclear corrections are usually applied by assuming 
that heavy target effects are the same in neutrino and 
charged lepton induced reactions. However, there is 
{\em a priori} 
no reason why this  should be the case. Hence, it is important  
to investigate the role played by shadowing in neutrino
reactions {\em before} concluding that the two structure functions 
are {\em really} different in the small $x$-region.   

(ii) Apart from  heavy target corrections,  uncertainties in the 
strange and anti strange quark 
distributions can also effect the comparison of the two structure functions.   

\section{Comparison of neutrino and muon structure functions} 

Comparisons of structure functions measured
in neutrino deep-inelastic scattering with those
measured in charged lepton deep-inelastic scattering are based on
the interpretation of these structure functions in terms of
parton distribution functions in the quark parton model.
Assuming the validity of charge symmetry and neglecting
the contributions from charm quarks,
the structure functions $F_2^{\nu N_0}(x,Q^2)$ and
$F_2^{\ell N_0}(x,Q^2)$ on iso-scalar targets ($N_0$)
are given by the following expressions:
\begin{eqnarray}
  F_2^{\ell N_0}(x,Q^2) & =& \frac{5}{18} x[ u(x) + \bar u(x)
 +d(x) +\bar d(x) + \frac{2}{5} (s(x) + \bar s(x))] \\
  F_2^{\nu N_0} (x, Q^2) &=& x[u(x)+ \bar u(x) +d(x) +\bar d(x)
     + 2 s(x)].
\end{eqnarray}
Thus, they can be  related to each other  by
\begin{equation}
  F_2^{\ell N_0}(x,Q^2)= \frac{5}{18}
F_2^{\nu N_0 }(x,Q^2) - \frac{3x[s(x)+\bar s(x)]+5x[s(x)-\bar s(x)]}
{18}.
\label{eq:1}
\end{equation}
This means that, once the charged lepton and neutrino structure functions
and the strange quark distributions are known, one can test the validity
of this relation, or one can use the above relation to extract the
strange quark distribution from the measured structure functions.

A quantity which is very sensitive to any deviations of the two structure 
functions from each other is the ``charge ratio'' defined as 
\begin{equation}
 R_c(x,Q^2)\equiv \frac{F_2^{\mu N_0}(x,Q^2)}{\frac{5}{18}
 F_2^{\nu N_0}(x,Q^2) -\frac{x(s(x)+\bar s(x))}{6} }
\approx 1 - \frac{s(x)-\bar s(x)}{Q_s(x)}, 
\label{Rc}
\end{equation}
where I introduced the notation 
$Q_s(x) \equiv \sum_{q=u,d,s} [q(x)+\bar q(x)] -
3(s(x)+\bar{s}(x))/5$. 
Assuming charge symmetry and that heavy target 
correction are under control and 
expanding the ``charge ratio'' in first order 
in small quantities  
it is clear from Eq.\ref{Rc} that any deviation from unity is proportional 
to the difference between the strange and anti strange quark 
distributions. The possibility of different strange and anti strange 
quark distributions should be explored later. Let us focus on 
nuclear corrections, first.

\begin{figure}[t]
\centering{ \hbox{ \hspace{1.cm}
\psfig{figure=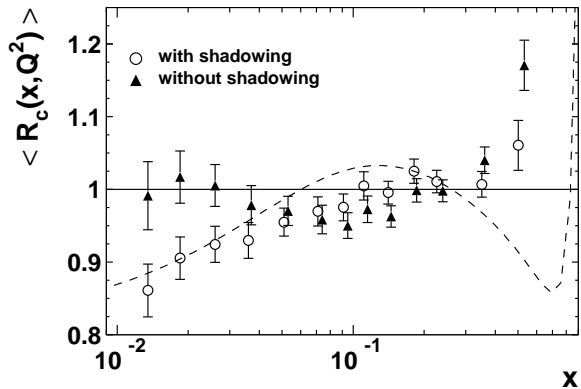,height=2.4in} \hspace{1.cm}
  }}
\caption{The ``charge ratio'' $R_c$ of Eq.\ \protect\ref{Rc}
         as a function of $x$ calculated using
         the CCFR \protect\cite{CCFR} data for neutrino and
         NMC \protect\cite{NMC} data for muon
         structure functions. The data have been integrated
         over the overlapping kinematical regions
         and  have been corrected for heavy target effects
         using a parametrization (dashed line) for heavy target corrections
         extracted from charged lepton scattering. The result is
         shown as open circles.
         The ratio  obtained without heavy target corrections
         is shown as solid triangles.
         Statistical and systematic errors are added in
         quadrature.   }
\label{fig1}
\end{figure}

In order to demonstrate the importance of the nuclear corrections 
we calculated the ``charge ratio'' using the neutrino and charged 
lepton structure functions 
extracted by the CCFR and NMC Collaborations, respectively. 
In correcting the neutrino structure function for 
nuclear effects we made two different  assumptions and 
integrated the data over $Q^2$ for fixed $x$ in 
the overlapping kinematical regions of the two experiments. 
The result is shown in Fig.\ref{fig1}.  
The solid triangles stand for the ``charge ratio'' assuming that 
there is no nuclear corrections in neutrino 
scattering and the open dots are the results assuming 
that the nuclear corrections are the {\it same} in neutrino and 
charged lepton scattering. The parametrization of the nuclear 
corrections in charged lepton scattering 
 used in correcting the data is shown as 
dashed line.  We see that while, {\it without } nuclear corrections,  
the two structure functions would be compatible with each other 
there is a considerable discrepancy between them in the small 
$x$ region  when nuclear corrections in neutrino reactions 
are comparable to those in charged lepton induced reactions. 

\section{Shadowing corrections}

In  calculating  the shadowing corrections we use a two-phase model
which has been  successfully applied to the description of  shadowing
 in charged-lepton deep inelastic scattering \cite{Badelek,Melni}.
This approach uses vector meson dominance (VMD)  to describe
the low-$Q^2$, virtual photon interactions, and Pomeron exchange
for the approximate scaling region. It is ideally suited
to describe the transition region between large  and
small $Q^2$.  This is the kinematic region where the largest
differences occur between the NMC and CCFR data sets.

In generalizing this approach
to weak currents we found that the essential differences in shadowing
between neutrino and charged lepton
deep inelastic scattering are:
(i) the axial-vector current is only partially conserved,
in contrast to vector currents \cite{Adler};  and (ii) the weak current
couples not only to vector but also to axial vector mesons
\cite{Stodolsky,VMD,Bell,Boris1}.

Partial conservation of the axial current (PCAC) requires
that the divergence of the axial current does not vanish
but is proportional to the pion field for $Q^2=0$. This
is Adler's theorem \cite{Adler},
which relates the neutrino cross section to the pion
cross section  on the same target for $Q^2=0$.
Thus, for low $Q^2\approx m_\pi^2$ shadowing in neutrino
scattering is determined by the absorption of  pions on the target.
For larger $Q^2$-values the contributions of vector and axial vector mesons
become important. The coupling of the weak current
to the vector and axial
vector mesons and that of the electro-magnetic current
to vector mesons are related to each other by the ``Weinberg sum rule''
$f_{\rho^+}^2=f_{a_1}^2=2f_{\rho^0}^2$.
Since the coupling of the vector (axial vector)
mesons to the weak current is  twice as large as  the coupling
to the electro-magnetic current but the structure function is 
larger by a factor of $\sim18/5$  in the neutrino
case, we expect that shadowing due to VMD in neutrino reactions is
roughly half of that in charged lepton scattering.
For larger $Q^2$-values, shadowing due to Pomeron exchange between the
projectile and two or more constituent nucleons dominates. Since
Pomeron-exchange models the interaction between partons
in different nucleons and the scattering of the $W$ takes
place on only one parton, this processes is of leading twist in
contrast to the VMD and pion contributions.
The coupling is given by the coupling
of the photon or $W$ to the quarks in the exchanged Pomeron.
It changes in the same way as the structure function does
in switching from neutrino to charged lepton
scattering. Thus, for large $Q^2$ values ($>10$ GeV$^2$) shadowing in both
cases should have approximately the same magnitude.

In summary, we expect no essential differences in shadowing between 
neutrino and muon scattering both for very small $Q^2$ and for very large 
$Q^2$-values. 
However, 
in the intermediate $Q^2$-region ($1<Q^2<10$ GeV$^2$), where VMD
is relatively important, we expect to see differences in shadowing
between neutrino and charge lepton scattering.
We recall, that this is precisely the region where
the discrepancy between CCFR and NMC is significant.
\begin{figure}[t]
\centering{ \hbox{ \hspace{1.cm}
\psfig{figure=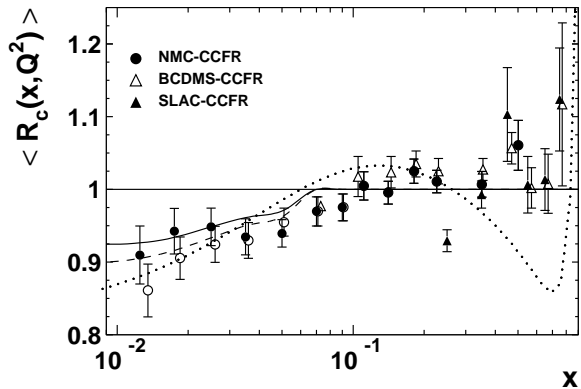,height=2.4in} \hspace{1.cm}
  }}
\caption{The charge ratio as a function of $x$  calculated using
         the CCFR \protect\cite{CCFR} data for neutrino and
         NMC \protect\cite{NMC}, SLAC \protect\cite{SLAC} 
        and BCDMS \protect\cite{BCDMS} data for muon
         induced structure functions. The data have been integrated
         above $Q^2=2.5$ GeV$^2$ over the overlapping kinematical regions and
         the statistic and systematical errors are added in
         quadrature.
         The heavy target corrections are calculated by using the
         ``two phase-model''
         in the shadowing region
         and a fit to the experimental data on nuclear shadowing
         in the non-shadowing region (black circles) and by using
         the $Q^2$ independent fit in the entire region (open circles).
         The ratio $R=F_2^{Fe}/F_2^D$ calculated for neutrino
         and for charged lepton scattering,
          is shown as solid  and dashed lines,
         respectively. They are calculated in the ``two phase model''
         and are averaged over the same $Q^2$-regions as the data.
         The $Q^2$ independent fit is represented by a
         dotted line.  }
\label{fig2}
\end{figure}

We calculated the shadowing corrections to the CCFR neutrino
data using the two-phase model of Ref. \cite{Badelek,Melni}.  
For details of the calculation see Ref.\cite{Boros1}. 
With the corrected CCFR data, we calculated the charge ratio $R_c$ of
Eq.\ \ref{Rc} between CCFR \cite{CCFR} and NMC \cite{NMC}, 
the CCFR and SLAC \cite{SLAC}, and 
CCFR and BCDMS \cite{BCDMS} data. 
The result is shown in
Fig.\ref{fig2}. The open 
circles show the charge ratio when heavy target shadowing corrections
from charged lepton reactions are applied to the neutrino data,
and the solid circles show the result when the neutrino shadowing
corrections from our two-phase model are applied. 
The ratio $R=F_2^{Fe}/F_2^D$ calculated for neutrino 
and for charged lepton scattering, 
is shown as solid  and dashed lines, respectively. 
 For $x \ge 0.1$, the
two shadowing corrections give essentially identical results.  At
small $x$, using the ``correct'' neutrino shadowing
corrections reduces the deviation of the charge ratio from unity.
Furthermore, the $Q^2$-dependence of the shadowing corrections plays
an important role in these corrections 
as can be seen from the solid, dashed and dotted lines in Fig.\ref{fig2}.  
Nevertheless, the charge 
ratio is still not compatible with one at small $x$. 

Our main conclusion is that 
properly accounting for shadowing corrections in the
neutrino structure function decreases, but does not resolve,
the low-$x$ discrepancy between the CCFR and the NMC data.

\section{Strange quark distribution} 

\begin{figure}[t]
\centering{ \hbox{ \hspace{1.cm}
\psfig{figure=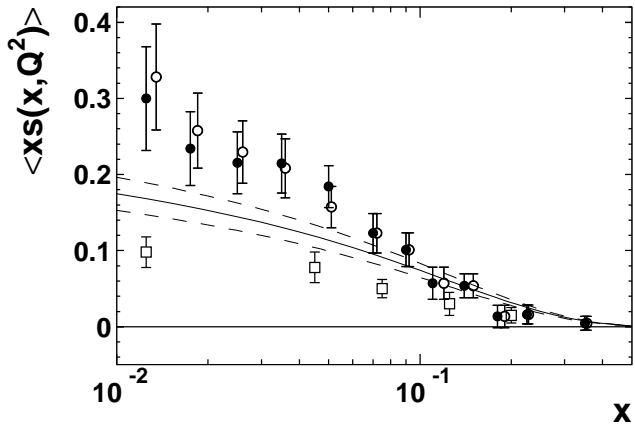,height=3.in} \hspace{1.cm}
}}
\vspace*{-1.cm} 
\caption{The  strange quark distribution extracted
         from the CCFR and NMC data assuming the validity of
        charge symmetry and $s(x)=\bar s(x)$.
        The data have been integrated over the overlapping
        $Q^2$ region to obtain better statistics.
        The solid (open)   circles stand for
        $5/6 F_2^{\nu}- 3 F_2^\mu$ using the two phase model
       (using the $Q^2$-independent parametrization)
        for the shadowing corrections.
        The open boxes stand for the LO CCFR determination of the strange
        quark density from dimuon production at $Q^2=4$ GeV$^2$
         \protect\cite{CCFRLO} .
        The solid line is the NLO CCFR determination at $Q^2=4$ GeV$^2$
          \protect\cite{CCFRstr}.
         The band around the NLO curve indicates the
        $\pm 1\sigma$ uncertainty in the distribution.
        }
\label{fig3}
\end{figure}
Shadowing corrections also influence the extraction of 
the strange quark distribution. 
Currently there are two viable
methods for the extraction of strange quark parton distributions.
The ``direct'' method utilizes charm-hadron production in neutrino
deep-inelastic scattering. The triggering signal for this process
is the measurement of opposite sign dimuons, one coming from the
lepton vertex, while the other comes from the semi-leptonic decay of
the charmed hadron \cite{CCFRLO,CCFRstr}. The other method is to obtain
the strange quark distribution by comparing charged lepton deep
inelastic scattering with neutrino deep inelastic scattering. In the
second case the strange quark distribution can be extracted from
the relation
\begin{equation}
 \frac{5}{6}F_2^{\nu N_0}(x,Q^2)-3F_2^{\mu N_0}(x,Q^2) =xs(x)+\frac{x}{3}
[s(x)-\bar s(x)].
\label{eq:ssbar}
\end{equation}
Eq.\ \ref{eq:ssbar} follows if we assume parton charge symmetry and
neglect charm quark contributions.
If one assumes that $s(x)=\bar s(x)$, the difference between the
neutrino and muon structure functions
measures the strange quark distribution in the nucleon.
However, it is found that 
the two methods for determining the strange quark
distribution are not compatible in the region of small $x$.
This conflict is also reflected in the fact that the
``charge ratio'' $R_c$ is different from one in this region using 
a strange quark distribution extracted from the di-muon 
experiment.

We converted the CCFR neutrino data on iron to deuteron data by
applying our shadowing corrections.  We then extracted
the strange quark distribution according to Eq.\ \ref{eq:ssbar}.
In order to get better statistics we integrated the structure functions
over the overlapping $Q^2$-regions, as before.
The result  is shown in Fig.\ref{fig3}, where the strange
quark distributions extracted with
the ``two-phase'' shadowing and the ``$Q^2$-independent''
shadowing corrections are  shown as black and open circles, respectively.
Statistical and systematic errors are added in quadrature.
The strange quark distribution
as determined by the CCFR Collaboration
in dimuon production using a LO analysis \cite{CCFRLO} is shown
as open boxes, while the distribution extracted in
NLO analysis \cite{CCFRstr}
from dimuon data is shown as a solid line. The band around the NLO curve
indicates the $\pm 1\sigma$ uncertainty.
Although the strange quark distribution obtained from the difference
between the neutrino and muon structure functions using the ``two phase''
model for shadowing is smaller in the small $x$-region than that obtained
by applying the $Q^2$-independent shadowing, both distributions
are incompatible with the strange quark distribution extracted from
dimuon production.

\begin{figure}[t]
\centering{ \hbox{ \hspace{1.cm}
\psfig{figure=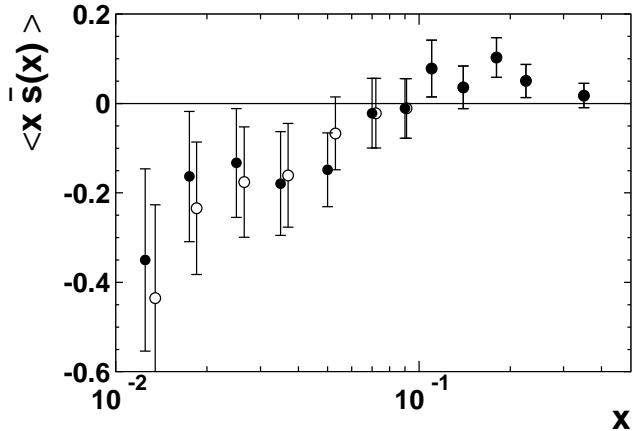,height=3.in} \hspace{1.cm}
}}
\vspace*{-1.cm} 
\caption{The (physically unacceptable)
 anti-strange quark distribution  extracted from the data
   assuming that the discrepancy between the muon and neutrino structure
   function is due to different strange quark and anti-strange quark
   distributions and that the strange quark distribution is given
   by that extracted from  di-muon experiments. }
\label{fig4}
\end{figure}

The remaining discrepancy could be attributed to {\it different}
strange and anti-strange quark distributions \cite{ST,Brodsky,sdiff,JT,HSS} 
in the nucleon. From Eqs.\ (\ref{Rc}) and (\ref{eq:ssbar}) and
Fig. \ref{fig3}, we see that
the difference $s(x)-\bar s(x)$ should be positive for small
$x$-values ($x<0.1$). This is in contradiction with
the analysis of Ref.
\cite{Brodsky} but agrees qualitatively with that in Ref. \cite{sdiff}.
Note in this connection that the experimentally determined
structure function, $F_2^{CCFR}$, is a flux weighted average of the neutrino
and anti neutrino structure functions \cite{CCFR}.
Since neutrino events dominate
over the anti neutrino events in  the event sample of the CCFR
experiment, it can be approximately regarded as neutrino
structure function.  
In Fig.\ref{fig4} we extract the strange antiquark distribution
vs.\ $x$ using Eq.\ (\ref{eq:ssbar}).  We use the experimental data
for the muon and neutrino structure functions (with our calculated
shadowing corrections), together with the strange quark distribution
measured in dimuon production.  Note that with this method
we obtain a {\it negative} strange antiquark distribution for
small $x$-values!  This means that the two experiments 
(di-muon production and structure functions measurements in muon and 
neutrino scattering) are incompatible with each other even if 
the anti strange quark density is completely unconstrained 
\cite{Boros2}.  
(The small admixture of anti neutrino events does not influence 
this conclusion \cite{Boros2}.)   
Thus, the entire discrepancy
cannot be attributed to  different $s(x)$ and $\bar s(x)$ 
distributions. 

\section*{Conclusions} 

In conclusion, we have carefully re-examined shadowing corrections to
the structure function $F_2^\nu$ in deep inelastic neutrino
scattering on an iron target. Although the shadowing corrections are not
as large as one would naively expect, they are still
sizable and similar to shadowing in charged lepton induced
reactions in the small-$x$ region. Taking neutrino
shadowing corrections into account properly
resolves part of the discrepancy between the
CCFR neutrino and the NMC muon data in the small $x$-region.
Neutrino shadowing corrections also remove part of
the corresponding discrepancy between the
two different determinations of the strange
quark densities.  However, the charge ratio $R_c$, of Eq.\
\ref{Rc}, still deviates from unity at small $x$.
Furthermore, the
data rules out the possibility that the discrepancy is entirely
due to the difference between the strange and anti-strange
quark distributions.
We are therefore forced to consider the possibility of a rather
uncomfortably large charge symmetry violation in the sea quark
distributions \cite{Lond,Sather,Lon98,Ben97,Ben98}.  
This is   discussed in Ref. \cite{Boros2} in more detail.

\section*{Acknowledgements} 
This research was performed in collaboration with 
J. T. Londergan and A. W. Thomas and was supported in part by 
the Australian Research Council and  the NSF under contract NSF-PHY/9722706.  
I would like to thank J. T. Londergan and A. W. Thomas 
for their stimulating and pleasant collaboration.

\end{document}